\documentclass{aa}
\usepackage{graphicx}
\begin{document}
\title{Strong gravitational lensing:
 Why no central black holes?}
\author{Da-Ming Chen}
\institute{National Astronomical Observatories, Chinese Academy of Sciences,
Beijing 100012, China}
\offprints{Da-Ming Chen, \email{cdm@class1.bao.ac.cn}}
\date{Received 12 July 2002 / Accepted 03 September 2002}
\abstract{We investigate how central black holes (BHs) in galactic
dark halos could affect strong gravitational lensing. The
distribution of integral lensing probability  with image
separations are calculated for quasars of redshift 1.5 by
foreground dark matter halos. The mass density of dark halos is
taken to be the Navarro-Frenk-White (NFW) profile such that, when
the mass of a halo is less than $10^{14} M_{\sun}$, its central
black holes or a bulge is included as a point mass. The
relationship between the masses $M_{\bullet}$ of supermassive
black holes and the total gravitational mass $M_{\mathrm{DM}}$ of
their host galaxy is adopted from the most recent literature. Only
a flat $\Lambda$CDM model is considered here. It is shown that,
while a single black hole for each galaxy contributes considerable
but not sufficient lensing probabilities at small image
separations compared with those without black holes, the bulges
(which are about $100$--$1000$ times larger in mass than a typical
black hole) would definitely contribute enough probability at
small image separations, although it gives too high probabilities
at large separation angles compared with lensing observations.
\keywords{gravitational lensing -- galaxies: black holes --
galaxies: bulge}}
\titlerunning{Why no central black holes?}
\authorrunning{D.-M. Chen}
\maketitle 

\section{Introduction}
Cold Dark Matter (CDM) has become the standard theory of
cosmological structure formation. The $\Lambda$CDM variant of CDM
with $\Omega_m=1-\Omega_{\Lambda}\approx 0.3$ appears to be in
good agreement with the available data on large scales (Primack
\cite{primack}). On smaller (sub-galactic) scales, there seem to
be various discrepancies, such: N-body CDM simulations which give
cuspy halos with divergent profiles towards the center (Navarro,
Frenk and White \cite{nfw96}, \cite{nfw97}, NFW hereafter); bar
stability in high surface brightness spiral galaxies which also
demands low-density cores; CDM models which yield an excess of
small scale structures; formation of disk galaxy angular momentum,
which is much too small in galaxy simulations. Issues that have
arisen on smaller scales have prompted people to propose a wide
variety of alternatives to CDM, such as warm dark matter (WDM) and
self-interacting dark matter (SIDM). Now that problems arise from
galaxy-size halos and centers of all dark matter halos,
high-resolution simulations and observations are the final
criterion. Recent highest-resolution simulations appear to be
consistent with NFW (Klypin, \cite{klypin}; Power et al.
\cite{power}) until scales smaller than about 1 kpc. Meanwhile, a
large set of high-resolution optical rotation curves has recently
been analyzed for low surface brightness (LSB) galaxies. One can
also conclude that the NFW profile is a good fit down to about 1
kpc. Although further simulations and observations, including
measurement of CO rotation curves (Bolatto et al., \cite{bolat}),
may help to clarify the nature of the dark matter, it now appears
that WDM and SIDM are both probably ruled out, while the
small-scale predictions of $\Lambda$CDM may be in better agreement
with the latest data than appeared to be the case as recently as a
year ago.

In addition to direct simulations and observations, gravitational
lensing provides another powerful probe of mass distribution in
the universe. Since mass within small scales only deflect light
rays slightly, it is difficult to extract mass information from a
single lensing event, and thus statistical gravitational lensing
is needed even for ``strong" gravitational lensing of small
halos(Turner, Ostriker \& Gott \cite{turner}; Narayan \& White
\cite{narayan}; Cen et al. \cite{cen}; Kochanek \cite{kochanek};
Wambsganss et al. \cite{wambs95}; Wambsganss, Cen, \& Ostriker
\cite{wambs98}; Porciani \& Madau \cite{porci}; Keeton \& Madau
\cite{keeton}). Li \& Ostriker (\cite{li}) first used the
semi-analytical approach to analyze gravitational lensing of
remote quasars by foreground dark halos in various cold dark
matter cosmologies. The mass function of dark halos they used is
alternatively given by singular isothermal sphere (SIS), the NFW
profile, or the generalized NFW profile. They found that none of
these models can completely explain the current observations: the
SIS models predict too many large splitting lenses, while the NFW
models predict too few small splitting lenses, so they proposed
that there must be at least two populations of halos in the
universe: small mass halos with a steep inner density slope and
large mass halos with a shallow inner density slope. The author
conclude that a combination of SIS and NFW halos can reasonably
reproduce the current observations. Similarly, Sarbu et al.
(\cite{sarbu}) investigated the statistics of gravitational lenses
in flat, low-density cosmological models with different cosmic
equations of state $\omega$. It was found that COBE-normalized
models with $\omega > -0.4$ produce too few arcsecond-scale lenses
in comparison with the JVAS/CLASS radio survey, a result that is
consistent with other observational constraints on $\omega$.

When attention is attracted to  alternatives of CDM dark matter
density profile at small scales, another kind of dark matter ---
super-massive black holes in the centers of most galactic halos is
forgotten or ignored in this case, although the idea of detecting
supermassive compact objects by their gravitational lensing
effects was proposed very early (Press \& Gunn \cite{press73},
Wilkinson et al. \cite{wilki}) and the lensing effects of
Schwarzschild black holes in the strong field regime have been
discussed in detail (e.g., Virbhadra \& Ellis \cite{virbh};
Frittelli, Kling \& Newman \cite{fritt}; Bozza et al.
\cite{bozza}). On the other hand, cosmological voids can form
directly after the collapse of extremely large wavelength
perturbations into low-density black holes or cosmological black
holes; such black holes can also be detected through their  weak
and strong lensing effects (Stornaiolo \cite{storn}). The
observational evidence presented so far suggests the ubiquity of
BHs in the nuclei of all bright galaxies, regardless of their
activity, and BH masses correlate with masses and luminosities of
the host spheroids and, more tightly, with stellar velocity
dispersions (Magorrian et al. \cite{magor}; Ferrarese \& Merritt
\cite{ferra}; Ravindranath et al. \cite{ravin}; Merritt \&
Ferrarese \cite{merria}, \cite{merrib}; Wandel \cite{wandel};
Sarzi et al. \cite{sarzi}). Most recent high-resolution
observational data gives $M_{\bullet}/M_\mathrm{bulge}\approx
10^{-3}$(Merritt \& Ferrarese \cite{merric}). Ferrarese
(\cite{ferra02}) further gave the relation between masses
$M_{\bullet}$ of supermassive black holes and the total
gravitational mass of the dark matter halo in which they
presumably formed
\begin{equation}
\frac{M_{\bullet}}{10^8M_{\sun}}\sim
0.046\left(\frac{M_\mathrm{DM}}{10^{12}M_{\sun}}\right)^{1.57}.
\label{bd}
\end{equation}
In this paper, we investigate the contributions of galactic
central black holes to lensing probabilities at small image
separations. Since $\Lambda$CDM cosmology and NFW profile are in
good agreement with the available data of structure formation on
almost all scales as mentioned above, we only chose these two
models respectively as cosmology and mass density function in our
calculations. We model the lenses as a population of dark matter
halos with an improved version of the Press-Schechter
(\cite{press74}, PS) mass distribution function, and central BHs
are considered for galaxy-size halos.

The paper is organized as follows: the lensing equation is given
in Sect.~\ref{s2}, lensing probabilities are calculated in
Sect.~\ref{s3}, and discussion and conclusions are provided in
Sect.~\ref{s4}.

\section{Lensing equation}\label{s2}
The NFW profile is
\begin{equation}
\rho_\mathrm{NFW}=\frac{\rho_\mathrm{s}r_\mathrm{s}^3}
{r(r+r_\mathrm{s})^2}
\end{equation}
where $\rho_\mathrm{s}$ and $r_\mathrm{s}$ are constants. We can
define the mass of a halo to be the mass within $r_{200}$ (which
is the radius of a sphere around a dark halo within which the
average mass density is $200$ times the critical mean mass density
of the universe),
\begin{equation}
M_\mathrm{DM}=4\pi\int^{r_{200}}_0\rho
r^2dr=4\pi\rho_\mathrm{s}r_\mathrm{s}^3f(c_1), \label{mdm}
\end{equation}
with $c_1=r_{200}/r_\mathrm{s}$ the concentration parameter, the
value of which is chosen to be $7/(1+z)$ (Bartelmann el al.
\cite{barte}). And
\begin{equation}
f(c_1)=\int^{c_1}_0\frac{x^2dx}{x(1+x)^2}=\ln(1+c_1)-\frac{c_1}{1+c_1}.
\end{equation}
In flat $\Lambda$CDM cosmology, the constants $\rho_\mathrm{s}$
and $r_\mathrm{s}$ can then be expressed as (Li \& Ostriker
\cite{li}),

\begin{equation}
\rho_\mathrm{s}=\rho_\mathrm{crit}\left[\Omega_\mathrm{m}(1+z)^3
+\Omega_{\Lambda}\right]\frac{200}{3}\frac{c_1^3}{f(c_1)},
\end{equation}

\begin{equation}
r_\mathrm{s}=\frac{1.626}{c_1}\frac{M_{15}^{1/3}}
{\left[\Omega_\mathrm{m}
(1+z)^3+\Omega_{\Lambda}\right]^{1/3}}h^{-1}\mathrm{Mpc}.
\label{rs}
\end{equation}
where $\rho_{crit}$ is the present value of the critical mass
density of the universe, and $M_{15}$ is the reduced mass of a
halo defined as
$M_{15}=M_\mathrm{DM}/(10^{15}\mathrm{h}^{-1}M_{\sun})$.

 The surface mass density for NFW profile is
\begin{eqnarray}
\Sigma_\mathrm{NFW}(x)&=&2\rho_\mathrm{s}r_\mathrm{s}\times\left\{
\begin{array}{ll}
\frac{\ln x-\sqrt{1-x^2}-\ln(1-\sqrt{1-x^2})}{(1-x^2)^{3/2}}, &
\ \ \ \ (x>1),\\
\frac{1}{3}, & \ \ \ \ (x=1), \\
\frac{\arcsin(1/x)+\sqrt{x^2-1}-\pi/2}{(x^2-1)^{3/2}}, & \ \ \ \
(0<x<1).
\end{array}. \right.
\end{eqnarray}
Where $x=|\vec{x}|$ and $\vec{x}=\vec{\xi}/r_\mathrm{s}$,
$\vec{\xi}$ is the position vector in the lens plane. The galactic
central black holes are assumed to be point masses, and we
consider fist there is only a single black hole with mass
$M_{\bullet}$ for each galaxy. So the surface mass density for
galactic halos each with a single central black hole can be
written as
\begin{equation}
\Sigma_\mathrm{galaxy}(\vec{x})=M_{\bullet}\delta^2(\vec{x})
+\Sigma_\mathrm{NFW}(\vec{x}),
\end{equation}
where $\delta^2(\vec{x})$ is the two dimensional Dirac-delta
function. The lensing equation with galactic central black holes
considered then is
\begin{equation}
y=x-\mu_\mathrm{s}\frac{f_{\mathrm{BH}}+g(x)}{x},\label{lq}
\end{equation}
where $y=|\vec{y}|$,
$\vec{\eta}=\vec{y}D_\mathrm{S}^\mathrm{A}/D^\mathrm{A}_\mathrm{L}$
is the position vector in the source plance, in which
$D_\mathrm{S}^\mathrm{A}$ and $D_\mathrm{L}^\mathrm{A}$ are
angular-diameter distances from the observer to the source and to
the lens respectively. And
\begin{equation}
\mu_\mathrm{s}=\frac{4\rho_\mathrm{s}r_\mathrm{s}}{\Sigma_\mathrm{cr}},
\end{equation}
where $\Sigma_\mathrm{cr}=(c^2/4\pi
G)(D_\mathrm{S}^\mathrm{A}/D_\mathrm{L}^\mathrm{A}
D_\mathrm{LS}^\mathrm{A})$
is the so called critical surface mass density, in which
$D_\mathrm{LS}^\mathrm{A}$ is the angular-diameter distance from
the lens to the source. And
\begin{eqnarray}
g(x)&=&\ln\frac{x}{2}+\left\{
\begin{array}{ll}
\frac{\arctan\sqrt{x^2-1}}{\sqrt{x^2-1}} & \ \ \ \ (x>1),\\
1 & \ \ \ \ (x=1), \\
\frac{\mathrm{arctanh}\sqrt{1-x^2}}{\sqrt{1-x^2}} & \ \ \ \
(0<x<1).
\end{array} \right.
\end{eqnarray}
In Eq.(\ref{lq}), the term \textbf{$f_{\mathrm{BH}}$} stands for
the contribution of a black hole, and by using Eq.(\ref{bd}) and
Eq.(\ref{mdm}), it has the form
\begin{equation}
f_{\mathrm{BH}}=2.78\times 10^{-4}f(c_1)M_{15}^{0.57}. \label{f}
\end{equation}
Since there is always more than one black holes in a bulge, and
thus the bulge itself can act like a black hole, we can treat a
bulge as a point mass in this paper as an extreme case. The mass
of a black hole correlates linearly with that of its host bulge as
$M_{\bullet}/M_\mathrm{bulge}\approx 10^{-3}$, so we can simply
multiply the term $f_{\mathrm{BH}}$ by $10^3$ to stand for the
contribution of a bulge. However, some light rays from the source
will definitely travel across the bulge, so there must exist a
kind of ``effective" black hole with mass larger than a single
``real" black hole but less than the bulge. In order to
investigate the tendency of image separations contributed by
different point mass, we can multiply $f_{\mathrm{BH}}$ by, for
example, $10^{2}$, etc.

\begin{figure}
\resizebox{\hsize}{!}{\includegraphics{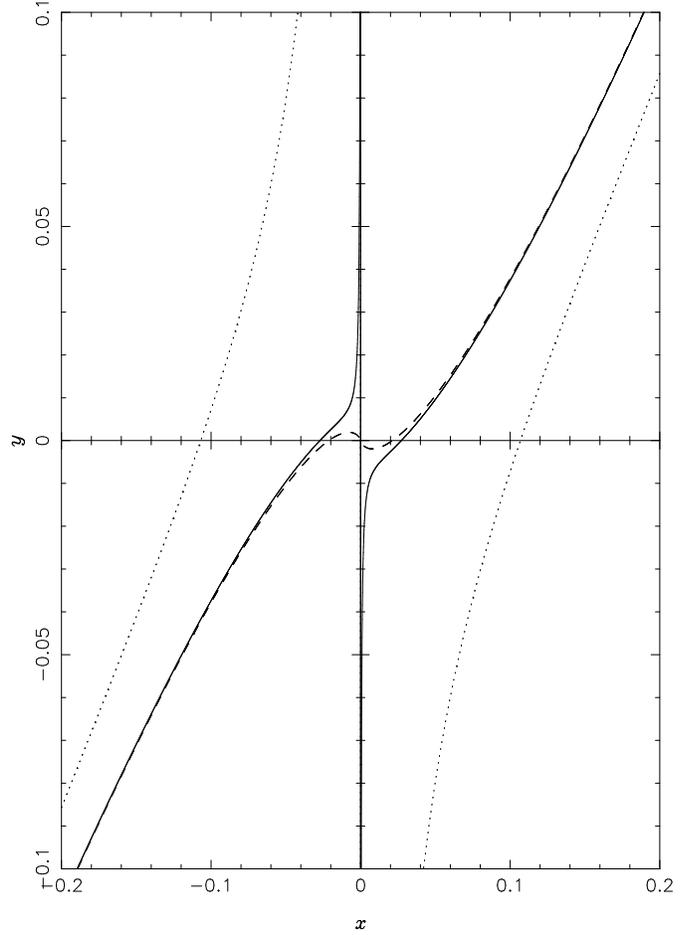}} \caption{Lensing
equations: dashed line stands for a NFW lens with no central black
holes; solid line stands for a lens with a NFW density profile
plus a single central black hole; dotted line stands for a lens of
NFW plus a bulge (treated as a point mass, an effective black
hole). } \label{fig1}
\end{figure}

The lensing equations for three cases are plotted in
Fig.~\ref{fig1} according to Eq.(\ref{lq}), where we have extended
$x$ and $y$ to their opposite values because of symmetry. The full
line, dashed line and dotted line, respectively, represent the NFW
lens with $f_{\mathrm{BH}}=0$, `NFW+BH' lens with
$f_{\mathrm{BH}}=2.78\times 10^{-4}f(c_1)M_{15}^{0.57}$ and
`NFW+bulge' with $f_{\mathrm{BH}}=2.78\times
10^{-2}f(c_1)M_{15}^{0.57}$, in which $\mu_{\mathrm{s}}=0.49$ and
$f(c_1)=0.91$.

We find that, as point masses, both a single central black hole
and a bulge can more often produce small separation images than
the case when no central black holes are considered. This result
will be further confirmed by the lensing probability given in next
section.

\section{Lensing probability}\label{s3}
We choose the most generally accepted values of the parameters for
flat $\Lambda$CDM cosmology, for which, with usual symbols, the
matter density parameter, vacuum energy density parameter and
Hubble constant are respectively: $\Omega_{\mathrm m}=0.3$,
$\Omega_{\Lambda}=0.7$, $h=0.75$.

The quasars of redshift $z_{\mathrm{s}}=1.5$ are lensed by
foreground CDM halos of galaxy clusters and galaxies, the lensing
probability with image separations larger than $\Delta\theta$ is
(Schneider, Ehlers, \& Falco \cite{schne})
\begin{equation}
P(>\Delta\theta)=\int^{z_{\mathrm{s}}}_0\frac{dD_{\mathrm{L}}(z)}
{dz}dz\int^{\infty}_0\bar{n}(M,z)\sigma(M,z)dM. \label{prob1}
\end{equation}
Where $D_{\mathrm{L}}(z)$ is proper distance from the observer to
the lens located at redshift $z$
\begin{equation}
D_{\mathrm{L}}(z)=\frac{c}{H_0}\int^z_0\frac{dz}{(1+z)
\sqrt{\Omega_{\mathrm{m}}(1+z)^3+\Omega_{\Lambda}}},
\end{equation}
here $c$ is the speed of light in vacuum and $H_0$ is the current
Hubble constant. The physical number density $\bar{n}(M,z)$ of
virialized dark halos of masses between $M$ and $M+dM$ is related
to the comoving number density $n(M,z)$ by
$\bar{n}(M,z)=n(M,z)(1+z)^3$; the latter was originally given by
Press \& Schechter (\cite{press74}), and the improved version is
\begin{equation}
n(M,z)dM=\frac{\rho_0}{M}f(M,z)dM,
\end{equation}
where $\rho_0$ is the current mean mass density of the universe,
and
\begin{equation}
f(M,z)=-\sqrt{\frac{2}{\pi}}\frac{\delta_c(z)}{M\Delta}
\frac{d\ln\Delta}{d\ln
M}\exp\left[-\frac{\delta_c^2(z)}{2\Delta^2} \label{ps}\right]
\end{equation}
is PS mass function. In Eq.(\ref{ps}) above, $\Delta^2(M)$ is the
present variance of the fluctuations in a sphere containing a mean
mass $M$,
\begin{equation}
\Delta^2(M)=\frac{1}{2\pi^2}\int^{\infty}_0P(k)
W^2(kr_{\mathrm{M}})k^2dk,
\end{equation}
where $P(k)$ is the power spectrum of density fluctuations,
$W(kr_{\mathrm{M}})$ is the Fourier transformation of a top-hat
window function
\begin{equation}
W(kr_{\mathrm{M}})=3\left[\frac{\sin(kr_{\mathrm{M}})}
{(kr_{\mathrm{M}})^3}-\frac{\cos(kr_{\mathrm{M}})}
{(kr_{\mathrm{M}})^2}\right],
\end{equation}
and
\begin{equation}
r_{\mathrm{M}}=\left(\frac{3M}{4\pi\rho_0}\right)^{1/3}.
\end{equation}
In Eq.(\ref{ps}), $\delta_c(z)$ is the over density threshold for
spherical collapse by redshift $z$ (Navarro, Frenk, \& White
\cite{nfw97}):
\begin{equation}
\delta_{c}(z)=\frac{1.68}{D(z)},
\end{equation}
where $D(z)$ is the linear growth function of density perturbation
(Carroll, \& Press \cite{carroll})
\begin{equation}
D(z)=\frac{g(\Omega(z))}{g(\Omega_{\mathrm{m}})(1+z)},
\end{equation}
in which
\begin{equation}
g(x)=\frac{5}{2}x\left(\frac{1}{70}+\frac{209x}{140}
-\frac{x^2}{140}+x^{4/7}\right)^{-1},
\end{equation}
and
\begin{equation}
\Omega(z)=\frac{\Omega_{\mathrm{m}}(1+z)^3}
{1-\Omega_{\mathrm{m}}+\Omega_{\mathrm{m}}(1+z)^3}.
\end{equation}
We use the fitting formulae for CDM power spectrum $P(k)$ given by
Eisenstein \& Hu (\cite{eisen})
\begin{equation}
P(k)=AkT^2(k),
\end{equation}
where $A$ is the amplitude normalized to
$\sigma_8=\Delta(r_{\mathrm{M}}=8h^{-1}\mathrm{Mpc})=0.95$, and
\begin{equation}
T=\frac{L}{L+Cq^2_{\mathrm{eff}}},
\end{equation}
with
\begin{equation}
L\equiv\ln(e+1.84q_{\mathrm{eff}}),
\end{equation}
\begin{equation}
q_{\mathrm{eff}}\equiv\frac{k}{\Omega_{\mathrm{m}}h^2{\mathrm{Mpc^{-1}}}},
\end{equation}
\begin{equation}
C\equiv14.4+\frac{325}{1+60.5q_{\mathrm{eff}}^{1.11}}.
\end{equation}

We need to know the cross-sections in Eq.(\ref{prob1}). Since we
are interested in the lensing probabilities with image separations
larger than a certain value $\Delta\theta$ (ranging from $0\sim
10$ arcseconds, for example), the cross-section is defined under
the condition that the multiple images can be created. For the
lenses with NFW profile, one can see from Fig.\ref{fig1} that
multiple images can be produced only if $|y|\leq y_{\mathrm{cr}}$,
where $y_{\mathrm{cr}}$ is the maximum value of $y$ when $x<0$,
which is determined by $dy/dx=0$ when $f_{\mathrm{BH}}=0$ in
Eq.(\ref{lq}). For galaxy-size halos, the mass of which is
confined to be less than $10^{14}M_{\sun}$ through out this paper,
a central black hole or a bulge as a point mass is considered (see
Eq.(\ref{lq})). In this case, multiple images will always exist:
when the source is close to the point caustic, i.e., when $y$ is
small, there are three images, two of which are within the
Einstein circle and the third one the outside which Einstein
circle; when $y$ is large enough, there are two images, the weaker
one is close to the center of the Einstein circle and the brighter
one locates outside of the Einstein circle. So another condition
is needed to define the cross-section, for which we use the
brightness ratio between the brighter and weaker images just
mentioned, and it is enough to set the ratio to be $10$.

The brightness ratio $r$ for the two images is just the ratio of
the corresponding absolute values of magnifications (Schneider,
Ehlers, \& Falco \cite{schne}),
\begin{equation}
r=\left|\frac{\mu_+}{\mu_-}\right|,
\end{equation}
where
\begin{equation}
\mu_+(y(x))=\left(\frac{y}{x}\frac{dy}{dx}\right)_{x>0},
\end{equation}
\begin{equation}
\mu_-(y(x))=\left(\frac{y}{x}\frac{dy}{dx}\right)_{x<0}.
\end{equation}
Once the source position $y_{\mathrm{cr}}$ is determined by
\begin{equation}
|\mu_+(y_{\mathrm{cr}})|=10|\mu_-(y_{\mathrm{cr}})|,
\end{equation}
the cross-section can be calculated, both with and without central
black holes, as
\begin{equation}
\sigma(M, z)=\pi
y_{\mathrm{cr}}^2r_{\mathrm{s}}^2\vartheta(M-M_{\mathrm{min}}),
\end{equation}
where $\vartheta(x)$ is a step function, and $M_{\mathrm{min}}$ is
determined by lower limit of image separation
\begin{equation}
\Delta\theta=\frac{r_{\mathrm{s}}\Delta
x}{D_{\mathrm{L}}^{\mathrm{A}}}\approx\frac{2x_0r_{\mathrm{s}}}
{D_{\mathrm{L}}^{\mathrm{A}}} \label{dtheta}
\end{equation}
and Eq.(\ref{rs}) as
\begin{equation}
M_{\mathrm{min}}=8.927\times
10^{-8}M_{15}\left(\Omega_{\mathrm{m}}(1+z)^3+\Omega_{\Lambda}\right)
\left(\frac{c_1D_{\mathrm{L}}^{\mathrm{A}}\Delta\theta}{x_0}\right)^3.
\end{equation}
In Eq.(\ref{dtheta}), we have approximated the image separation
$\Delta x$ to be $2x_0$, where $x_0$ is the positive zero position
of function $y(x)$ , both when $f_{\mathrm{BH}}=0$ (for NFW lens
only) and $f_{\mathrm{BH}}\neq 0$ (for galactic, NFW+BH/bulge
lenses) in Eq.(\ref{lq}), since image separation is insensitive to
the source position $y$ (Li \& Ostriker, \cite{li}).
\begin{figure}
\resizebox{\hsize}{!}{\includegraphics{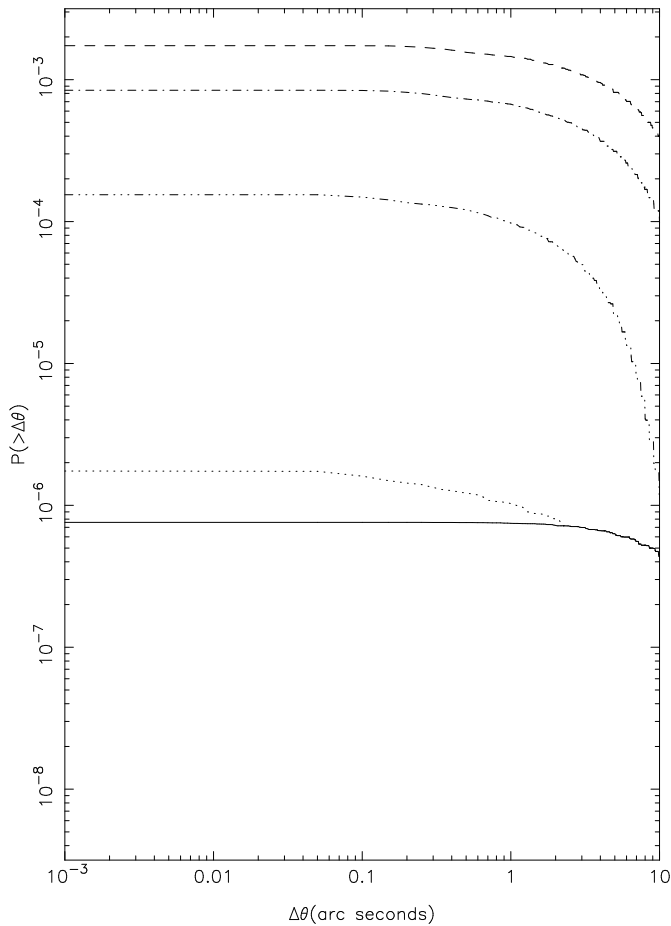}} \caption{Lensing
probability with image separations larger than $\Delta\theta$: the
full line is for lenses of NFW density profile with no central
black holes for halos at all scales, dotted line shows the case
when a single central black hole for each galaxy-size halo is
included. Other three lines show the cases when collectors of
central black holes in the bulge are treated as an effective black
hole. The dashed, dot-dash and dash-dot-dot-dot line, from top
downwards, show respectively, the mass of the effective black hole
at $1000$, $500$ and $100$ times that of a single `real' black
hole.} \label{fig2}
\end{figure}
We plot the lensing probability with image separations larger than
$\Delta\theta$ in Fig.~\ref{fig2}. In order to show the tendency
of contributions to the lensing probability for different fraction
of the bulge mass (the so called `effective' black hole), we take
the term $f_{\mathrm{BH}}$ in Eq.(\ref{lq}) to be
$f_{\mathrm{BH}}=2.78\times 10^{-1}f(c_1)M_{15}^{0.57}$,
$f_{\mathrm{BH}}=2.78\times(5.0\times 10^{-2})f(c_1)M_{15}^{0.57}$
and $f_{\mathrm{BH}}=2.78\times 10^{-2}f(c_1)M_{15}^{0.57}$; they
are represented by the first three lines from top down,
respectively. When central black holes or bulges are included and
treated as point masses, the mass of their host halos is confined
to be less than $10^{14}M_{\sun}$, because Eq.(\ref{bd}) strictly
applies in the range $10^6<M_{\bullet}<2\times 10^9M_{\sun}$ and
$10^{14}M_{\sun}$ is the upper-limit of galaxy mass (Ferrarese
2002).

\section{Discussion and conclusions}\label{s4}
Our numerical results for lensing probability with image
separations larger than $\Delta\theta$ in five different cases are
shown in Fig.~\ref{fig2}. In all cases, lensing probabilities keep
nearly constant until $\Delta\theta\sim 0.1$ arc seconds, and
obvious dropdown takes place at about $1$ arc second if central
black holes are included, which, of course, does not mean that the
main lensing events have image separations larger than $0.1$ arc
seconds. As a matter of fact, in the NFW case (without galactic
central black holes, the full line in Fig.~\ref{fig2}), the
lensing probability drops quite slowly in the whole range of image
separations: $\Delta\theta\sim$ 0---10 arc seconds; such a
tendency would extend even to $30$ arc seconds if it is plotted
beyond this range, which implies a uniform distribution of lensing
probability for its $log$ value among image separations. However,
note that in the single black hole case (dotted line), the lensing
probability drops to the same value of NFW at 2 arc seconds, which
gives the influence range of a single black hole. In the range of
$0\sim 0.1$ arc seconds, the lensing probability for the single
black hole case is about 3 times that for NFW. So, clearly, the
contributions from central black holes cannot be omitted, although
such contributions alone are indeed not enough to explain the
observational data. As we have pointed out, there is always more
than one black hole in a galactic bulge, and the collector of
black holes would make a bulge itself `act like' a black hole. On
the other hand, not all the mass of a bulge is concentrated in
black holes, so if we treat a whole bulge as an extreme black
hole, such a model would produce too many lenses at image
separations larger than 3 arc seconds compared with the JVAS/CLASS
radio survey. As mentioned above, we have sufficient reason to
tune the fraction of a bulge mass to produce a `right' profile of
lensing probability at larger image separations required by
observational data, but this `sufficient reason' seems not make us
produce sufficient lensing probabilities at smaller image
separations, as shown by the dash-dot line in Fig.~\ref{fig2}.

However, we can attribute sufficient lens events at small image
separations to galactic central black holes or the bulge. On the
one hand, since this paper focuses on whether galactic central
black holes would contribute considerably to the lensing
probability, we have not considered the effect of magnification
bias, which would increase the final result provided here at all
image separations. On the other hand, we have used an improved
version of the PS halo mass function but not the `best' version.
The shape of the mass function predicted by standard PS theory
(the improved version) is in reasonable agreement with what is
measured in numerical simulations of hierarchical clustering from
Gaussian initial conditions only for massive halos,; less massive
halos are more strongly clustered or less anti-biased than the
standard PS predicted. Sheth \& Tormen (\cite{st}, ST) proposed a
model that provides a reasonably good fit to the bias relation of
less massive haloes as well as to that of massive halos. Note that
central black holes are only found in galactic bulges,; ST's
correction for mass function in the range of less massive halos
would definitely change the lensing probability  discussed in this
paper. Also note that one of the two images produced by a galactic
central black hole is close to the lens center and very faint;
however, VLBI experiments can detect its existence (Hirabayashi
\cite{hirab}; Ulvestad \cite{ulves}), and further radio lensing
surveys would have the ability to identify high flux density ratio
of the two images. How and to what extent lensing magnification
bias, flux ratio and modified PS mass function may change the
final result will be discussed in another paper.

\begin{acknowledgements}
I thank Professor Xiang-Ping Wu for his original idea for this
paper and the anonymous referee for useful comments and
suggestions. This work was supported by the National Natural
Science Foundation of China.
\end{acknowledgements}

\end{document}